# Evaluating the Accuracy of the Labeling System in Web of Science for the Sustainable Development Goals

Yu Zhao[*], Li Li[**] and Zhesi Shen[**]

[*] *zhaoyu.cugb@gmail.com*
0009-0006-5554-359X
School of Earth Sciences and Resources, China University of Geosciences, China

[**] *lili2020@mail.las.ac.cn; shenzhs@mail.las.ac.cn*
0000-0001-8326-3620; 0000-0001-8414-7912
National Science Library, Chinese Academy of Sciences, China

**Abstract**

Monitoring and fostering research aligned with the Sustainable Development Goals (SDGs) is crucial for formulating evidence-based policies, identifying best practices, and promoting global collaboration. The key step is developing a labeling system to map research publications to their related SDGs. The SDGs labeling system integrated in Web of Science (WoS), which assigns citation topics instead of individual publication to SDGs, has emerged as a promising tool. However we still lack of a comprehensive evaluation of the performance of WoS labeling system. By comparing with the Bergon approach, we systematically assessed the relatedness between citation topics and SDGs. Our analysis identified 15% of topics showing low relatedness to their assigned SDGs at a 1% threshold. Notably, SDGs such as '11 Cities', '07 Energy', and '13 Climate' exhibited higher percentages of low related topics. In addition, we revealed that certain topics are significantly underrepresented in their relevant SDGs, particularly for '02 Hunger', '12 Consumption', and '15 Land'. This study underscores the critical need for continual refinement and validation of SDGs labeling systems in WoS.

## 1. Introduction

The United Nations' Sustainable Development Goals (SDGs) represent a transformative vision for global progress, encompassing a wide array of social, economic, and environmental dimensions(UN, 2015). Science is one of the essential mechanisms to realize the SDGs(Allen et al., 2019; Mishra et al., 2024; Nature, 2021; Yeh et al., 2022). As the international community strives to achieve these ambitious goals by 2030, monitoring the research landscape that contributes to the SDGs has become a critical endeavour. Effective monitoring not only aids in gauging the impact of scientific work but also steers future research towards areas of pressing need(Jha et al., 2020; Mishra et al., 2024; Yamaguchi et al., 2023).

Bibliometrics and science mapping analysis provide a quantitative framework for tracking the structure, dynamics, and impact of scientific output and its influence on SDGs(Chen, 2017; Garechana et al., 2012; Körfgen et al., 2018; Raman, Kumar Nair, et al., 2023; Raman, Nair, et al., 2022; Raman, Nair, et al., 2023; Raman, Subramaniam, et al., 2022). One of the key challenges in this pursuit is the accurate identification and classification of research outputs that align with the SDGs(Armitage et al., 2020; Morales-Hernandez et al., 2022; Purnell, 2022). Given the expansive body of scientific literature, the multifaceted nature of the SDGs(Bennich et al., 2020; Landhuis, 2016; Sianes et al., 2022) and lack of vocabulary knowledge or semantic ambiguity of each SDG(Bascur et al., 2023), developing a robust labeling system is essential for precise tracking and meaningful analysis.

The Clarivate™ (formerly Thomson Reuters, favoured by the top 400 educational institutions from over 150 nations) plays a pivotal role in this regard through its topic-level SDGs labeling system (Clarivate, 2024; Garcia, 2022) both in the Web of Science database (WoS) and the InCites Benchmarking & Analytics platform (Incites). This system is designed to automatically categorize research publications into topics based on their citation relationships(Traag et al., 2019), thereby assigning SDGs labels to topics based on their relevance.

The accuracy and effectiveness of this system are paramount, as they are essential for equipping researchers, policymakers, and other stakeholders with the tools to effectively navigate the intricate relationship between scientific inquiry and SDGs. However, concerns have been raised about that, particularly in relation to the methodology of the topic identification which is based on a combination of bibliometric analysis and manual curation. The potential shortcomings of the system could lead to misalignment in the representation of research contributions to the SDGs, thereby undermining the integrity of monitoring efforts.

In this study, we perform a comparative analysis between WoS SDGs labeling system and Bergen topic-approach (BTA), one of the most used keyword-based queries easily redirecting to the WoS website with one-click, to identify questionable topics that require a reconsideration of their SDGs label.

The structure of this paper is as follows: Section 2 will delve into some background information on the SDGs labeling/mapping systems. Section 3 will outline the methodology and data used to assess the performance of the WoS SDGs labeling system. Section 4 will detail the findings of the study, highlighting the extent of concordance between the WoS SDGs labelling system and the BTA. Finally, Section 5 will discuss the implications of the findings, potential limitations, and propose recommendations for enhancing the accuracy and utility of SDGs labeling systems in scientometric analysis.

## 2. Literature Review

*2.1. Importance of SDGs labeling/mapping*

The potential for leveraging knowledge to foster SDGs holds promise, yet it faces significant hurdles. First of all, a significant portion of research produced by committed scholars often goes unused, relegated to libraries instead of propelling societal progress(Clark et al., 2016). The transition toward sustainability demands not just an increase in knowledge production, but a transformation in how knowledge is made accessible, applied, and actionable for the betterment of society. SDGs Mapping will help governments, civil society organizations, the private sector and individuals easily find research output related to specific SDGs.

Secondly, as Bali Swain (2017) pointed out, goals need to be quantifiable and measurable. Taking idea of scientometrics, SDGs mapping helps in defining clear indicators for each goal, which allows for the collection of academic data that can be used to measure progress in SDGs-related science from large-scale data sources.

The last but not the least, from a long-term vision and a systematic view, scientists in scientometrics are expected to provide policymakers, research funders and evaluation systems a roadmap to explore the structure and evolution of science(Fortunato et al., 2018) and help with three priority areas: removing roadblocks to progress; identifying transformation pathways; and improving governance(Malekpour et al., 2023). For examples, the misalignment or the inconclusive relationship between SDGs challenges and research prioritization will block the road to progress(Confraria et al., 2024). SDGs Mapping can support more evidence to prioritize which goals are most pressing and require immediate attention, allowing for the allocation of resources and efforts where they are most needed.

*2.2 The SDGs mapping approach of scientific literature*

Various labeling systems have been developed to map research outputs to the SDGs, operating as content-based classification systems, citation-based classification systems or the mix of the above two. Figure 1 illustrated the major labeling systems.

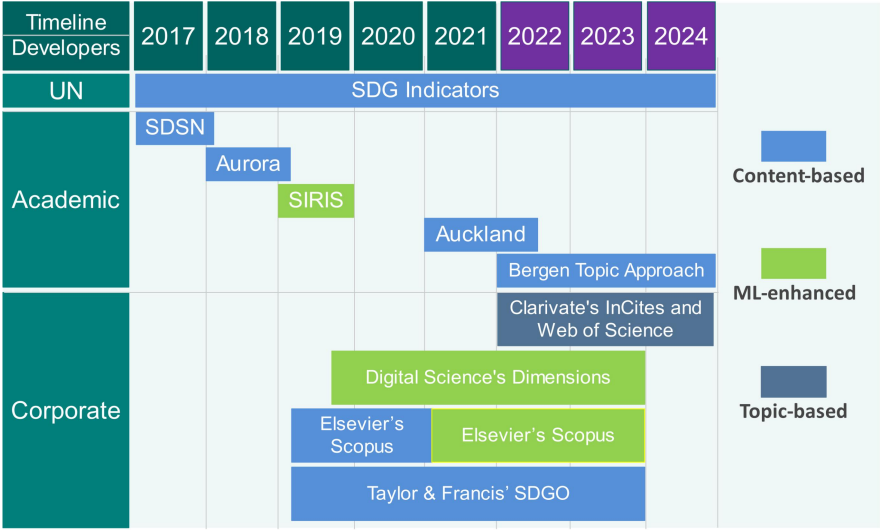

Figure.1. The SDGs mapping/labeling systems.

Content-based classification systems, such as string matching and supervised machine learning, analyze the intrinsic text features of documents. String matching, which was initially employed for SDGs tagging, provides the flexibility to adapt to regional characteristics and linguistic variations, yet its effectiveness can be compromised by the lack of vocabulary knowledge or semantic ambiguity. The 2017 webinar "Practical Approaches to Mapping University Contributions to the SDGs" marked a significant step forward by compiling the first list of SDGs keywords. Since then, a variety of string retrieval methods have been implemented, including those by Auckland, Aurora, SIRIS, Elsevier, SDGO, SDSN, and BTA. Supervised machine learning, introduced by Dimensions in 2019, leverages trained models to classify texts, offering benefits such as automation and reduced bias, albeit with limitations tied to the training data.

Topic-based topic classification methods leverage the citation networks within scientific literature to infer thematic connections. These methods offer a unique perspective that complements content-based approaches by focusing on the relationships among documents through citations. Only WoS has embraced this method, leveraging the citation analysis for a more comprehensive understanding of the relationship between research literature and the SDGs. The advantage of this method lies in its ability to capture the dynamic ecosystem of scientific ideas, where new scholarly contributions are linked to existing literature through citations. The disadvantage of this method lies particularly in relation to emerging fields or topics with limited citation history.

*2.3 From labeling single paper to labeling topics with SDGs*

In 2020, WoS has introduced a new research area classification system called "Citation Topics" , offering a novel way to analyze and refine research topics based on citation patterns. The Citation Topics is a three-level hierarchy of macro, meso and micro-level topics, with each topic being a cluster of papers that are densely citing each other. The Citation Topics are created based on nearly 70 million documents within WoS using the community dectection approach(Traag et al., 2019). Citation Topics includes 10 broad clusters, 326 meso-clusters, and 2444 micro-clusters as shown in Fig.2.

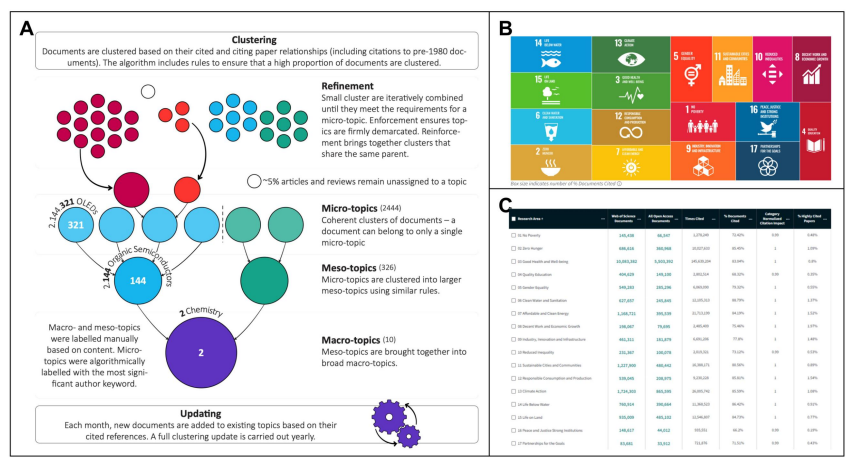

Figure 2. Snapshot of Citation topics: a three-level hierarchy of macro, meso, and micro topics

SDGs in WoS are based on Citation Topics, which leveraging the quantitative bibliometric analysis and the qualitative insights provided by experts's judgment. Since the release of SDGs labeling system in WoS, numerous scholars have harnessed this tool for identifying research articles associated with the SDGs and conducting scientometric research, which underscores the system's versatility and relevance. For instance, Rafael Repiso et al. conducted a comprehensive bibliometric analysis examining Spanish universities' research output related to the SDGs from 2017 to 2021. In another study, Hassam Bin Waseem et al. investigated the connections between social capital in flood risk reduction and the SDGs.

*2.4 Comparison among different classification systems*

Different labeling systems may generate diverse results. The evaluation of accuracy, consistency and disparity of labeling systems is another important problem. In the realm of evaluating the accuracy of SDGs labeling systems, two principal methodologies have been recognized: qualitative and quantitative, with the quantitative method being more common. A scalable solution in quantitative comparison includes benchmarking the labeling system against established classification systems, validating performance with precision, recall, and F1 scores, and using machine learning algorithms for automatic classification and accuracy assessment. Purnell (Purnell, 2022) compares four methods for identifying SDGs 13-related publications, revealing low overlap and highlighting the need for transparent documentation and careful method selection due to the subjective nature of search strategy and data source selection. Wang et al compares the intersection between search results from the Auckland system and WoS system(Wang et al., 2023).

## 3. Data and Methods

3.1 Data

Our evaluation is grounded in data extracted from the WoS databases including SCIE, SSCI and ESCI. The data was extracted on March 1, 2024, covering a time span from 2015 to 2023. This period aligns well with the United Nations' timeline for the SDGs, which were adopted in 2015 to be achieved by 2030.

Table 1. Basic statistics of the assignment of citation topics to SDGs in WoS.

|  | WoS |
| --- | --- |
| No. of all citation topics | 2,434 |
| No. of publications | 19,982,991 |
| No. of SDGs for this study | 10 |
| No. of citation topic-SDGs assignment pairs (within the 10 SDGs) | 1,807 |
| No. of SDG-related citation topics (within the 10 SDGs) | 1,680 |
| Max. no. of SDGs per citation topic (within the 10 SDGs) | 3 |
| Avg. no. of SDGs per citation topic (within the 10 SDGs) | 0.7 |

The dataset encompassed about 19 millions publications and 2,434 corresponding citation topics. This study primarily focused on 10 out of the 17 SDGs, including 01 Poverty, 02 Hunger, 03 Health, 04 Education, 07 Energy, 11 Cities, 12 Consumption, 13 Climate, 14 Water, and 15 Land. Through detailed analysis, it was found that 1,807 citation topics were successfully linked to one or more of these 10 SDGs. Notably, the maximum number of SDGs assigned to a single citation topic was 3, indicating a diverse range of thematic connections. On average, each citation topic was associated with approximately 0.7 of the 10 SDGs considered in this study.[1]

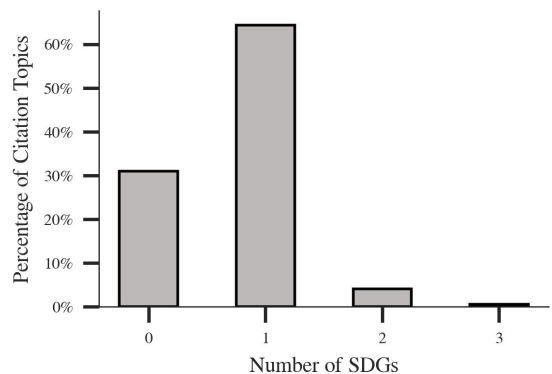

Figure 4. Number of SDGs per citation topic linked within 2434 citation topics

3.2 Methodology

In this study, we employed a string matching method to evaluate the accuracy of SDGs labeling systems in WoS. We selected the BTA as our benchmark due to its successful classification of publications related to the 10 SDGs and its widespread usage in query construction(Armitage et al., 2020; Purnell, 2022; Wulff et al., 2024). BTA has a twin method called Bergen action-approach (BAA), which comprehensively classifies research themes and specific actions related to SDGs. These approaches can be accessed at https://zenodo.org/records/10210818, providing direct query links to the WoS website.

The 'relatedness' $r_{i,c}$ between a citation topic $i$ and a SDGs $c$ was calculated as the proportion of papers whinin a citation topic matching the SDGs search strings, that is

$$r_{i,c} = \frac{n_{i,c}}{t_i}$$

where $n_{i,c}$ is the total number of publications retrieved by the searching string of SDGs $c$ within citation topic $i$, and $t_i$ is the total number of publications of citation topic $i$.

---

[1] It should be mentioned that citation topics with publication number < 100 will not be included in the presentation of the results of our analysis. Our string matching approach may not be sufficiently reliable for these topics. WoS has only one topics with publication number < 100, accounting for 0.04% of the total number of WoS topics. Taking a further look at this topic in WoS database, namely 7.70.1880 Falling Film, it turns out that only 1 paper was labeled with this topic and this topic has been deleted by Incites database.

Two criteria were established to identify potentially mislabeled citation topics based on theme relevance between papers in a specific citation topics and their assigned SDGs categories. **Criterion I**: This criterion targets citation topics that lack theme relevance to their assigned SDGs, suggesting that the classification might be questionable if $r_{i,c}$ is relatively low ($r_{i,c} \leq alpha$).

**Criterion II**: This criterion identifies citation topics that have strong theme relevance to SDGs which they are not assigned, indicating a possible error in the classification system if a citation topic is not categorized with a SDGs with which it has significant papers that match certain predefined strings for that SDGs ($r_{i,c} \leq beta$).

Threshold values for the criteria were set to determine what constitutes highly relevant to or irrelevant to SDGs according to the statistics on the proportion of the current 'SDG-related' citation topics or the current 'non SDG-related' citation topics. For Criterion I, thresholds of 0.1%, 1% and 3% were used, while for Criterion II, thresholds of 15%, 20% and 25% were applied.

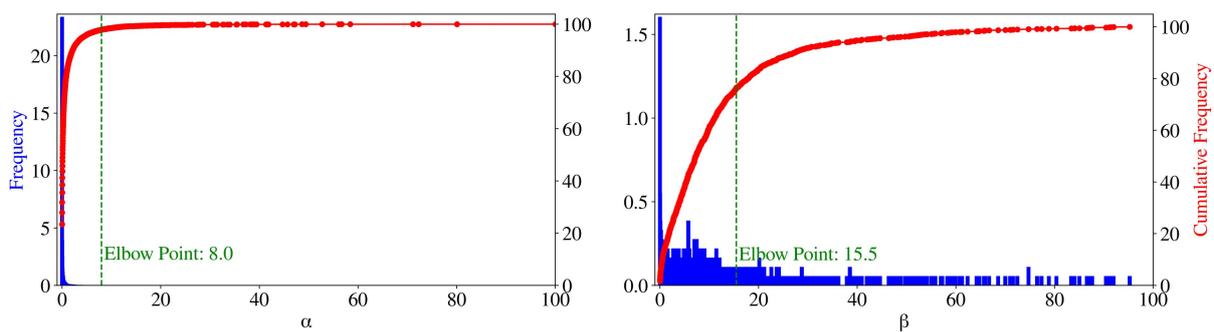

Figure 3.The distribution of the 'relatedness' between a citation topic and a SDG

## 4. Results

This section presents the performance of SDGs labeling system of WoS. Sections 5.1 and 5.2 provide the results obtained using Criteria I and II, respectively. Section 5.3 reports the results obtained by combining Criteria I and II. Section 5.4 presents an in-depth analysis for '11 Cities' and '07 Energy'.

**4.1 Criterion I: Citation topics with low relatedness to their assigned SDGs categories**

A citation topic satisfies Criterion I if it is assigned to a certain SDGs while the percentage of publications in this citation topic retrieved by the action searching strings or the topic searching strings of Bergen Approach belonging to the same category is relatively small. More precisely, a citation topic satisfies Criterion I if it is assigned to a certain SDGs c even though $r_{i,c}$ is below a certain threshold α. We use three values for the parameter α, i.e., 0.1%, 1%, and 3%. By using multiple parameter values, we get insight into the sensitivity of our results to the choice of the parameter value.

Table 2 Summary of the results from Criterion I

| Threshold α | No. of citation topic-SDG assignments (% of all citation topic-SDGs assignments) |
|---|---|
| ≤ 0.1 | 74 (4%) |
| ≤ 1 | 263 (15%) |
| ≤ 3 | 488 (27%) |

Table 2 reveals the number of citation topic-assigned SDGs pairs under different threshold values. At the threshold of 0.1%, 74 pairs, representing 4% of all topic-SDGs pairs, were identified. As can be expected, the trend indicates a proportional increase in both the number of citation topics and their associations with SDGs as the threshold value rises.

Table 3. Citation Topics satisfying Criterion I (Topic searching strings; α = 1%) for each SDGs

| SDG | No. of citation topics Labeled with this SDG | No. of citation topics with $r_{i,c} \leq 1\%$ |
|---|---|---|
| 11 Cities | 135 | 75 (56%) |
| 07 Energy | 74 | 30 (41%) |
| 13 Climate | 136 | 53 (39%) |
| 14 Water | 37 | 6 (16%) |
| 12 Consumption | 45 | 7 (16%) |
| 04 Education | 73 | 11 (15%) |
| 01 Poverty | 27 | 3 (11%) |
| 03 Health | 1132 | 72 (6%) |
| 02 Hunger | 66 | 3 (5%) |
| 15 Land | 81 | 3 (4%) |

We further investigated the problematic pairs for each SDGs. Table 3 shows the number of citation topics labeled with each SDG, along with the number and percentage of citation topics meeting the Criterion I at a threshold of 1%. Citation Topics related to '11 Cities', '07 Energy', and '13 Climate' exhibit higher percentages meeting the 1% threshold, emphasizing more attentions we may pay on these SDGs.

To facilitate the examination of the relationships among citation topics within the same SDG, we constructed a science map. In Fig.5, each point represents a citation topic, with the relative size of the points indicating the number of papers within that topic. The proximity of points reflects the strength of citation relationships between topics, with closer topics signifying a higher level of similarity and interdependence between them. We hypothesize that points representing more similar topics should be categorized under the same SDGs. We can see that there exists a pattern in the distribution of topics, where the upper left side of the figure demonstrates a stronger association with '11 Cities', while the lower right side shows weaker associations. This clustering pattern implies that there may exist systematic bias at macro-level fields.

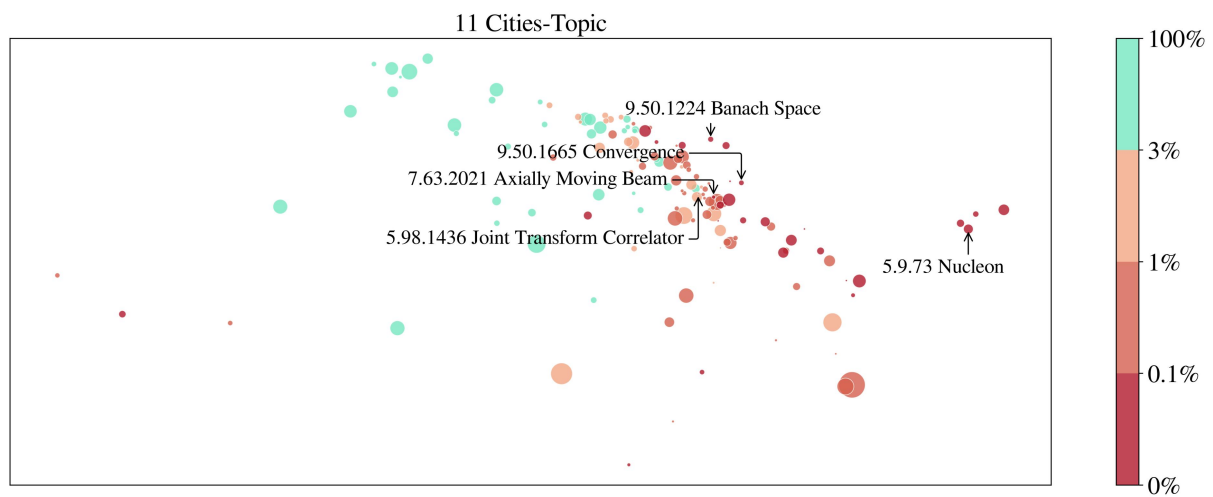

**Figure 5. Science map of Citation topics labeled with SDGs 11-Cities.** Each dot reprents a Citation Topic with size representing $t_i$ and color representing $r_{i,c}$. The distance among dots represent the similarity among citation topics, with more similar topics locating more closely.

By grouping all the micro-level citation topics into macro-level fields, Figure 6 reports the distribution of citation topics meeting the Criterion I at a threshold of 1% based on Topic searching strings across different SDGs and their macro-level field. The color intensity in the heatmap can indicate the percentage of citation topics falling under each field within a specific SDGs  For instance, for '03 Health', the majority of problematic citation topics are from Chemistry (50.0%), followed by Physics (19.4%). In contrast, '11 Cities' shows a diverse distribution, with problematic topics spreading across various macro fields like Engineering & Materials Science (38.7%) and Electrical Engineering, Electronics & Computing (25.3%) .

When focusing on the marco fields, we can see that fields such as Physics, Chemistry and Mathematics appear to be more susceptible to labeling errors across SDGs.

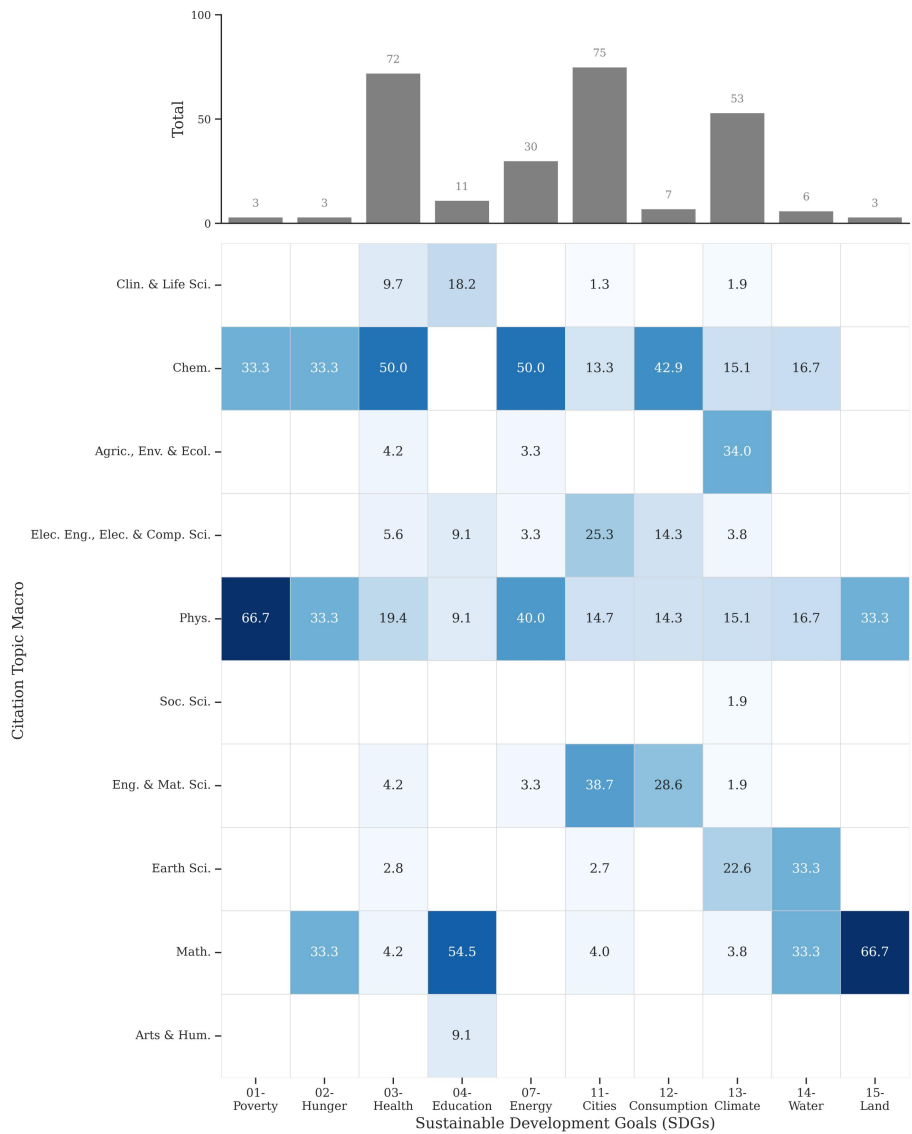

Figure.6. Fraction of Citation Topics satisfying Criterion I across fields. (α = 1%)

## 4.2 Criterion Ⅱ: Citation topics with high relatedness to SDGs which they are not assigned

Table 4 illustrates the outcomes of using three distinct thresholds ($\beta = 15\%, 20\%, 25\%$) to pinpoint citation topics lacking SDGs labeling. At a threshold of 20% and with the Topic searching strings, 87 citation topic-SDGs assignments were identified. As the threshold values increase, there is a noticeable decline in the count of non-SDGs citation topics identified, along with a reduction in the percentage of non-SDGs assignments.

**Table 4. Summary of the results from Criterion Ⅱ**

| Threshold $\beta$ | No. of citation topic-SDG assignments (% of all citation topic-SDGs assignments) |
|---|---|
| ≥ 15 | 157 |
| ≥ 20 | 87 |
| ≥ 25 | 57 |

**Table 5. Citation Topics satisfying Criterion Ⅱ for each SDG($\beta = 15\%$).**

| SDG | No. of citation topics labeled without this SDG | No. of citation topics with $r_{i,c} \geq 15\%$ |
|---|---|---|
| 02 Hunger | 2367 | 75 |
| 12 Consumption | 2388 | 34 |
| 15 Land | 2352 | 20 |
| 03 Health | 1301 | 8 |
| 01 Poverty | 2406 | 6 |
| 11 Cities | 2298 | 5 |
| 14 Water | 2396 | 5 |
| 07 Energy | 2359 | 3 |
| 13 Climate | 2297 | 1 |
| 04 Education | 2360 | |

Table 5 shows the number of citation topics meeting the Criterion Ⅱ for each SDGs at the threshold of 15%. SDGs such as '02 Hunger', '12 Consumption', and '15 Land' have more citation topics meeting the 15% threshold. When focusing on the macro fields, Social Sciences appear to be more susceptible to labeling errors, as shown in Fig.7.

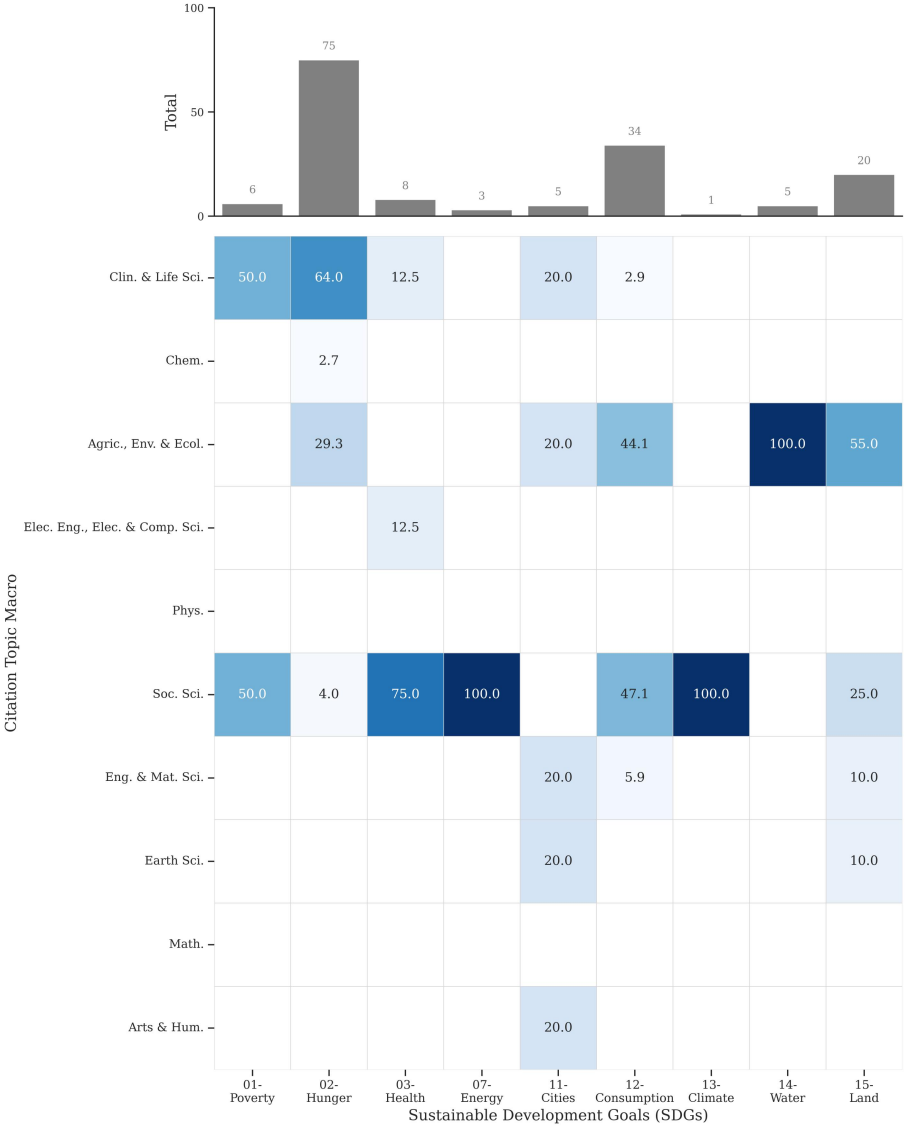

Figure 7. Fields satisfying Criterion Ⅱ (β = 15%).

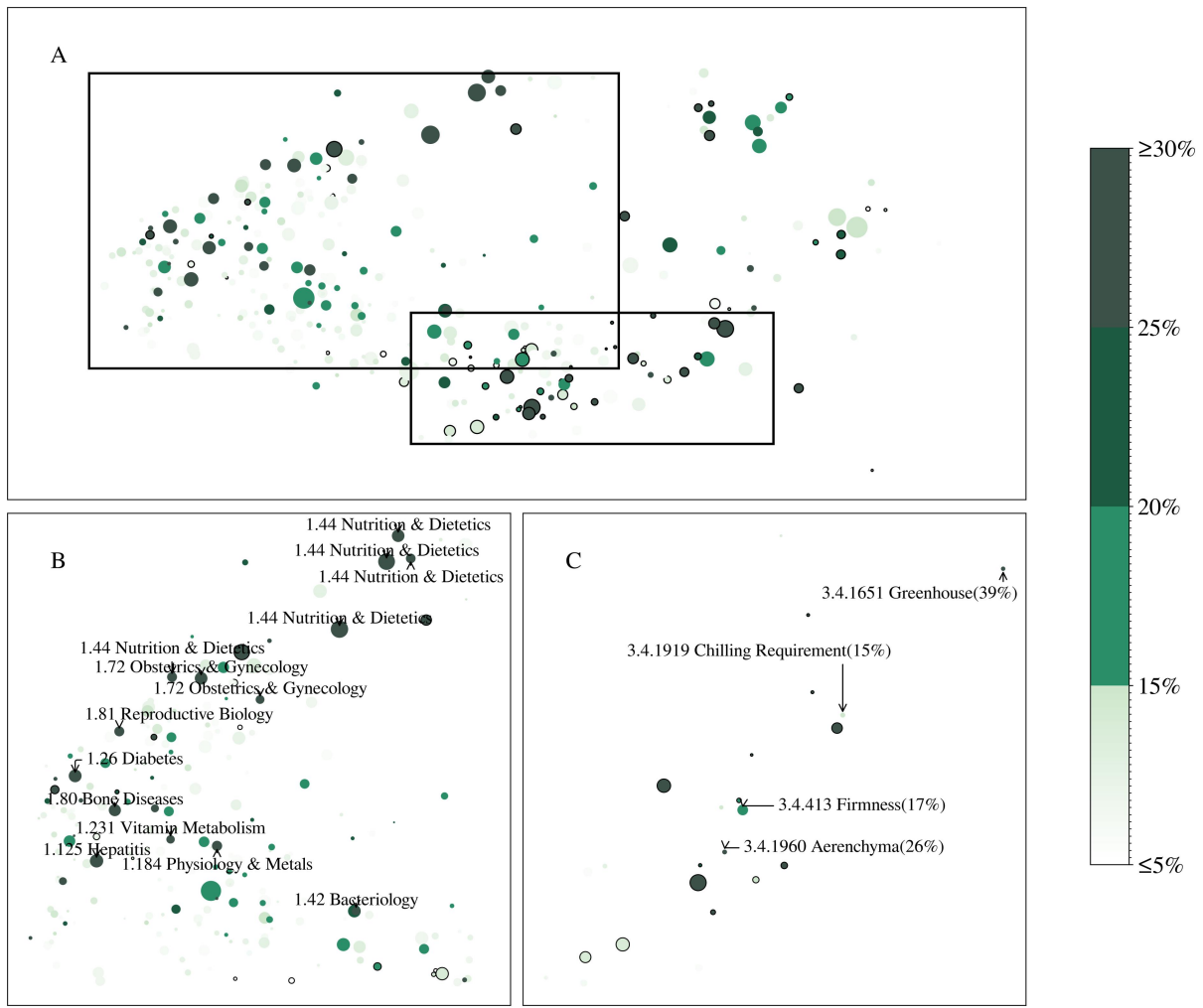

**Figure 8. Science map of citation topics related to SDG-02 Hunger.** (A) Distribution of citation topics related to 02 Hunger with color encoding the value of $r_{i,c}$. Dots with black boarder edges are topics labeled with 02 Hunger by WoS. (B) The zoom in of the upper left box in A. The meso-level topic name of dots with $r_{i,c}>25\%$ and $t_i > 10,000$ are highlighted. (C) The zoom in of the lower right box in A. Topics belonging to the meso-level topic-3.4 Crop Science-are presented.

As shown in Fig.7, SDG-02 Hunger has the largest number of highly related but unlabeled citation topics. To explore the patterns of unlabeled topics related to SDG-02, we presents the science map of citation topics in Fig.8(A). Two interesting areas can be observed. In the upper left black box area, there are many topics with $r_{i,c}>25\%$ but not labeled by WoS(the zoom in is shown in Fig.8(B)). Majorities of the topics are belonging to the meso-level topic-1.44 Nutrition & Dietetics implying that the related medical research may be ignored for SDG-02 Hunger in WoS labeling system. The another interesting area is the lower right black box area, in which the topics are mainly belonging to 3.4 Crop Science. As shown in the zoom in figure in Fig.8(C), there are a total of 25 citation topics. Among these, 13 are labeled as SDGs 02, with their $r_{i,c}$ all exceeding 15%. Additionally, among the 12 unlabeled topics, 4 topics having $r_{i,c}$ exceeding 15% and locating closely to the 13 labeled topics implying the internal inconsistency within the same meso-level topic.

### 4.3 Combine Criterion I and Criterion II

By combining the results of Criterion I ($\alpha = 1\%$) and Criterion II ($\beta = 15\%$), we revealed interesting insights into the potential reclassification of several citation topics. Notably, as shown in Table 7, citation topics-Liquid Chromatography and Brown Adipose Tissue- currently having low $r_{i,c}$ to their currently assgined SDGs but both having high $r_{i,c}$ to the unlabeled '02 Hunger'.

The potential relationship between these two citation topics with '02 Hunger'—rather than '11 Cities' or '03 Health'—can be attributed to their underlying connections to food security and nutritional aspects. Liquid chromatography is pivotal in food analysis and quality control, which are crucial for addressing hunger and malnutrition. Similarly, Brown adipose tissue is associated with metabolic health and energy balance, intersecting with the themes of hunger and food availability. Understanding and enhancing metabolic health can contribute to better nutritional strategies and energy utilization, which are essential in the fight against hunger.

Table 7 Citation topics satisfy Criterion I ($\alpha = 1\%$) and Criterion II ($\beta = 15\%$).

| Citation Topic Micro | Citation Topic Meso | Current labeled SDG | $r_{i,c}$ | Potential related SDGs | $r_{i,c}$ | Searching strings |
|---|---|---|---|---|---|---|
| 2.244.2352 Liquid Chromatography | 2.244 | 11 Cities | 0.2 | 02 Hunger | 37.2 | BTA |
| 1.26.1419 Brown Adipose | 1.26 | 03 Health | 0.9 | 02 Hunger | 20.0 | BAA |

| Tissue | | | | | | |
|---|---|---|---|---|---|---|

## 5. Discussion and Conclusion

The evaluation of SDGs classification systems within the WoS using keyword-based method has yielded significant insights into the accuracy and reliability of these systems. By employing the Bergen Approach as a benchmark, we were able to systematically assess the relatedness between citation topics and SDGs and identify potentially mislabeled citation topics, thereby highlighting areas for improvement in SDGs mapping.

Our analysis revealed that a substantial number of citation topics exhibit low relatedness to their assigned SDGs, as evidenced by Criterion I. Specifically, at a threshold of 1%, 263 citation topics (15% of all topics) were identified as potentially misclassified. This finding underscores the need for more stringent criteria and refined search strings to enhance the accuracy of SDGs assignments. Notably, SDGs such as 'SDGs 11 Cities', 'SDGs 13 Climate', and 'SDGs 07 Energy' were found to have higher percentages of misclassified topics, indicating that these categories may require more focused attention and potentially revised classification methodologies.

Criterion II further reinforced the presence of classification inconsistencies by identifying citation topics with high relatedness to SDGs to which they were not assigned. At a threshold of 20%, 87 citation topic-SDGs assignments were identified, suggesting that certain topics are significantly underrepresented in their relevant SDGs categories. This misalignment was particularly pronounced for SDGs like 'SDGs 02 Hunger', 'SDGs 12 Consumption', and 'SDGs 15 Land', where a notable number of citation topics met the 15% frequency threshold for relatedness but were not classified under these SDGs.

Combining the results from Criteria I and II provided a comprehensive overview of the potential reclassifications needed. Topics such as Liquid Chromatography and Brown Adipose Tissue demonstrated strong associations with SDGs 02, despite being classified under different SDGs. This combined analysis highlights the importance of a holistic approach to SDGs classification, where multiple criteria and thresholds are considered to ensure a robust and accurate system.

This study underscores the critical need for continual refinement and validation of SDGs classification systems within academic databases like WoS. The string matching method, particularly when using comprehensive search strings as in the Bergen Approach, proves to be a valuable tool in identifying misclassifications and enhancing the accuracy of SDGs assignments.

Our findings indicate that a significant portion of citation topics may be either misclassified or underrepresented in their relevant SDGs categories. Addressing these inconsistencies is essential for ensuring that the classification system accurately reflects the thematic connections between research outputs and SDGs. This, in turn, will facilitate more effective tracking of progress towards the United Nations' SDGs and support informed decision-making in research policy and funding.

Future work should focus on refining the search strings and thresholds used in the classification process, as well as exploring automated and machine learning-based approaches to further enhance accuracy. Additionally, expanding the analysis to include more SDGs and labeling systems will provide a more comprehensive understanding of the classification system's performance and areas for improvement. This work includes 10 SDGs provided by Bergon, and will benefit by comparing other labeling systems. A recent work compared 6 labeling systems based on topic coverage(Li et. al. 2024) showing more consistency among systems for several specific SDGs. Ultimately, a more accurate and reliable SDGs labeling system will better serve the research community and contribute to the global effort in achieving the SDGs by 2030.

**Open science practices**
Due to the nature of the research data used in this paper, which includes bibliographic information such as titles, abstracts, and keywords, access to the Web of Science (WoS) database is restricted by a contract with Clarivate that prohibits the redistribution of their data. Researchers interested in conducting their analysis with the raw data will need to obtain it through a paid subscription to Clarivate. The statistical results of our study will be made available on Figshare (https://figshare.com/), ensuring transparency and accessibility to the research findings. Additionally, the Bergen approach utilized for this study is accessed from https://zenodo.org/records/7241690.

**Acknowledgments**

This paper utilizes GPT technology for language polishing of the manuscript.

**Author contributions**

**Yu Zhao**: Writing – original draft, Formal analysis, Data curation, Methodology, Visualization, Conceptualization.

**Li Li**: Writing – review & editing, Formal analysis, Conceptualization.

**Zhesi Shen**: Writing – review & editing, Formal analysis, Supervision, Methodology, Conceptualization.

**Competing interests**

The authors declare that they have no conflict of interest.

**Funding information**